\documentclass{article}

\input{tcilatex}

\begin{document}

\title{Quantum-Statistical Computation}
\author{Giuseppe\ Castagnoli\thanks{%
Information Technology Division, Elsag spa, 16154 Genova, Italy} and David
Ritz Finkelstein\thanks{%
School of Physics, Georgia Institute of Technology, Atlanta, GA 30332, USA}}
\date{\today }
\maketitle

\begin{abstract}
Systems of spin 1, such as triplet pairs of spin-1/2 fermions (like
orthohydrogen nuclei) make useful three-terminal elements for quantum
computation, and when interconnected by qubit equality relations are
universal for quantum computation. This is an instance of \emph{%
quantum-statistical computation}: some of the logical relations of the
problem are satisfied identically in virtue of quantum statistics, which
takes no time. We show heuristically that quantum-statistical ground-mode
computation is substantially faster than pure ground-mode computation when
the ground mode is reached by annealing.
\end{abstract}

\section{Introduction}

\subsection{Premise}

To-day quantum computation is mostly algorithmical: a stored program
controls a time-varying Hamiltonian that sends a set of qubits -- two-valued
quantum variables -- through a sequence of unitary transformations effecting
steps in a computation. The quantum speed-up depends essentially on
maintaining coherent parallel computation paths during the entire
computation, until the final measurement.

The well-known fragility of such a coherence may make algorithmic quantum
computation impractical.

\emph{Quantum ground-mode computation} could be an interesting alternative.
One defines a time-independent quantum Hamiltonian proportional to the
amount of ``frustration'' (violation) of the logical relations of the
problem to be solved. Then any ground mode solves the problem.

An incoherent superposition of ground modes also solves the problem. This
should greatly reduce the problem of decoherence.

In quantum annealing computation, one form of quantum ground-mode
compuytation, the quantum computer is brought to minimum energy by a
suitable coupling with a heat bath of programmed temperature. In quantum
adiabatic computation (Farhi et al.), another form, the computer is prepared
in a minimum of a simplified Hamiltonian, which is then adiabatically
deformed into the problem Hamiltonian, leaving the network at minimum energy.

To be sure, quantum ground-mode computation is suspected of being
mathematically and physically intractable. At any rate, its computer
simulations are lengthy and seriously limit research on this method; and
since the numbers of energy levels, energy traps (for annealing
computation), and energy-level crossings or proximities (for adiabatic
computation), can grow exponentially with problem size, ground-mode
computation alone may not yield any speed-up.

We ameliorate this problem here. We implement part of the Boolean relations
as quantum symmetries expressing particle indistinguishability, using a
quantum-statistical three-terminal network element. Satisfying such
relations does not take time. There is a quantum speed-up for annealing
computation in the sense that the speed of relaxation of a hard problem
becomes comparable to that of an easy problem, which algorithmically can be
solved in polynomial time by successively eliminating variables.
Implementing part of the Boolean relations through quantum symmetries due to
particle indistinguishability should give a speed-up over the case where all
relations are implemented through an energy function.

\subsection{Outline}

We express the problem to be solved in terms of the
computationally-universal relations $q_{x}+q_{y}+q_{z}=1$, called the
(sum-1) triode, and $q_{x}=q_{y}$, called the wire. Here the $q$ are Boolean
variables and $+$ denotes arithmetical sum. Quantum-statistical computation
implements the wires through Hamiltonian terms, and the triodes through
quantum statistics (see also Castagnoli 1998, Castagnoli et al. 1998,
Castagnoli \& Monti 1999).

We associate each wire $q_{m}=q_{n}$ with a Hamiltonian term $%
(q_{m}-q_{n})^2 $ whose two-fold degenerate ground modes satisfy the wire.
We associate each triode with a spin pair that satisfies the relation
identically in virtue of particle statistics, without time development. The
associated Hamiltonian is 0. Satisfying relations implemented by statistics
does not take time or energy.

We show this here for quantum-statistical annealing (qusa) computation, by
means of a special representation of the computation process.

The effective Hamiltonian $H=H_{w}+H_{r}(t)$ consists of a Hamiltonian $%
H_{w} $ for the wires of the network, and a time-dependent heat-bath
coupling $H_{r}$ to relax the network to its zero point. There is no triode
Hamiltonian.

To estimate the solution time, we introduce a certain non-symmetrized
comparison Hamiltonian $H^{\prime }$ that also includes a coupling to a heat
bath and yields the actual network Hamiltonian $H$ when statistics --
symmetrization -- is imposed. $H^{\prime }$ describes the relaxation process
of an easy network, where all the Boolean constraints implemented by
statistics are removed. The easy network may be realized in principle by
replacing each pair of identical-indistinguishable fermions forming a triode
by a pair of distinguishable (e.g. non-identical)\ fermions. But when we
continuously project the process governed by $H^{\prime }$ on the Hilbert
space satisfying the statistics, we recover the relaxation process of the
actual network.

In this way we show that the solution time for the actual network is
comparable with that of the easy network.

Although the interplay between relaxation and statistics proposed is a
well-defined physical effect, the computation model based on it still dwells
in the same abstract conceptual realm as other current literature on quantum
ground-mode computation (see for example Farhi et al., among others). Our
model demonstrates that a new form of quantum speed-up is possible in
principle, leaving implementation problems for the future.

\section{The Boolean problem}

By a triode with qubits $q_{x},q_{y},q_{z}$ we mean the relation $%
q_{x}+q_{y}+q_{z}=1,$ where $+$ denotes arithmetical sum. This is the
negation of the POR (partial OR) relation used by Boole and algebraicized by
C. S. Peirce [Finkelstein 1996]. It is a partial NOR (or partial Scheffer
stroke) relation and can be written as $q_z = q_x$ PNOR $q_y$ .

We consider a Boolean triode network, one consisting only of $T$ triodes, $%
Q=3T$ qubits, and $W$ wires between qubits.

The triode is satisfied by just the values 
\begin{equation}
\frame{$%
\begin{array}{ccc}
q_{x} & q_{y} & q_{z} \\ \hline
0 & 0 & 1 \\ 
0 & 1 & 0 \\ 
1 & 0 & 0%
\end{array}%
$}
\end{equation}%
No value of $q_{x}$ is defined for $q_{y}=q_{z}=1$; and similarly for $q_{y}$
and $q_{z}$. This triode is therefore not a gate, since it does not define
an input-output function relating its variables. We may call this triode a
partial gate, since it defines a partial function; in fact, three.

A toy triode network is 
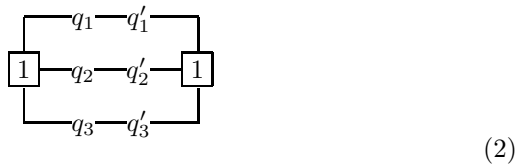
\begin{equation}  \label{eq:TOY}
\begin{picture}(200,60) \put(86,53){\line(0,-1){13}}
\put(86,50){$\rule[3pt]{18.0pt}{.5pt}q_1 \rule[3pt]{12.0pt}{.5pt}
q_1'\rule[3pt]{18.0pt}{.5pt}$} \put(152,53){\line(0,-1){13}}
\put(80,30){$\framebox{1}\rule[3pt]{12.0pt}{.5pt}q_2
\rule[3pt]{12.0pt}{.5pt} q_2'\rule[3pt]{12.0pt}{.5pt}\framebox{1}$}
\put(86,13){\line(0,1){13}} \put(86,10) {$\rule[3pt]{18.0pt}{.5pt}q_3
\rule[3pt]{12.0pt}{.5pt} q_3'\rule[3pt]{18.0pt}{.5pt}$}
\put(152,13){\line(0,1){13}} \end{picture}
\end{equation}%
with six qubits $q_{1},q_{2},q_{3},q_{1}^{\prime },q_{2}^{\prime
},q_{3}^{\prime }$, two triodes $q_{1}+q_{2}+q_{3}=1$ , $q_{1}^{\prime
}+q_{2}^{\prime }+q_{3}^{\prime }=1$ and three wires $q_{1}=q_{1}^{\prime
}\;,\;q_{2}=q_{2}^{\prime }\;,\;q_{3}=q_{3}^{\prime }$.

The problem is to solve the network assuming that there is at least one
solution. This problem has practically the same difficulty as checking
whether the network is satisfiable. It is the ``exact cover'' problem
addressed in quantum adiabatic computation (Farhi et al. 2001), and is
NP-complete.

\section{The network model}

\subsection{The qubit}

The qubits used in quantum computation differ from the bits of classical
computation in that a qubit $q$ has variables that do not commute with $q$,
and modes that are quantum superpositions of the $q=0$ and $q=1$ modes. Qusa
computation exploits the quantum nature of the qubit still further.

\subsection{The sum-1 triode}

We model the triode with a triplet proton pair identically fulfilling the
triode relation $q_{x}+q_{y}+q_{z}=1$ in virtue of statistics, as follows.

A quantum spin 1/2 provides three anticommuting two-valued variables $\sigma
_{x},\sigma _{y},\sigma _{z}$, each taking two values $\pm 1$, and subject
to the relation $\sigma _{x}\sigma _{y}\sigma _{z}=i$.

A hydrogen molecule has a triplet of ground modes and an excited singlet
mode. Let ${\frac{1}{2}}\sigma_{1},{\frac{1}{2}}\sigma_{2}$ be the
respective spin vectors of protons 1 and 2 (in units with $\hbar=1$). Then $%
s={\frac{1}{2}}(\sigma_{1}+\sigma_{2})$ represents the total spin angular
momentum.

Two independent spin vectors $\sigma _{1}$ and $\sigma _{2}$ define a total
spin $s=(\sigma _{1}+\sigma _{2})/2$ with $s_{z}=\pm 1,0$, and three
commuting binary variables $q_{x}=1-s_{x}^{2}$, $q_{y}=1-s_{y}^{2}$, $%
q_{z}=1-s_{z}^{2}$ subject to the relation $q_{x}+q_{y}+q_{z}=1$ mod 2.
These spins thus constitute a gate --  implementing the weaker relation $%
q_{x}+q_{y}+q_{z}=1$ mod 2 -- among the three qubits: 
\begin{equation}
\frame{$%
\begin{array}{ccc}
q_{x} & q_{y} & q_{z} \\ \hline
0 & 0 & 1 \\ 
0 & 1 & 0 \\ 
1 & 0 & 0 \\ 
1 & 1 & 1%
\end{array}%
$}  \label{eq:EQU}
\end{equation}%
Define the spin quantum number $S$ by 
\begin{equation}
s^{2}=S(S+1)  \label{eq:S}
\end{equation}%
as usual. The fourth line in this table is the singlet mode $S=0$. This gate
is the (non-universal) complemented-XOR gate. Since each variable is $1$
(true) when and only when the other two are equal, it may also be called the
EQUALS (or EQU) gate.

EQU is a functional relation. It may be solved for any of its qubits, say $%
q_{x}$: 
\begin{equation}
q_{x}=q_{y}+q_{z}+1\mbox{ mod }2\/.
\end{equation}

Networks of triodes and wires are universal for Boolean computation; it is
easy to construct NOT from two triodes and NOR from three triodes, with
several wires.

Networks of EQU gates and wires are not universal. A Boolean network made of
just EQU gates and wires is a system of modulo-2 arithmetical equations that
is quickly solvable.

In particular there is always the solution where all qubits are 1 and all
triodes are singlet.

When $q_z=0$ the total spin is ``up-or-down,'' which is doubly degenerate.
When $q_z=1$ it is ``not up-or-down,'' which is also doubly degenerate. The
qubits $q_x$ and $q_y$ have similar meanings relative to the $x$ and $y$
axes.

In the singlet mode, $q_{x}=q_{y}=q_{z}=1$.

If we restrict the system to its triplet mode $S=1$, the three commuting
qubits $(q_x, q_y, q_z)$ obey the PNOR relation 
\begin{equation}
q_x+q_y+q_z=1
\end{equation}%
by (\ref{eq:S}). Then ``not up-or-down" becomes (approximately) ``sideways"
and is non-degenerate. "Up-or-down" remains doubly degenerate.

In the comparison network constructed later we drop the restriction to the
triode relation. All triodes are replaced by EQU gates of truth table (3).

\subsection{The error metric}

We define an ``error metric'' measuring the distance of the network from a
solution. From now on we use the following notation. We index the network
wires with ${\omega }=1,\dots ,W$ and the triodes with $\tau =1,...,T$. For
each wire ${\omega }$, with terminal qubits $q( {\omega },0) $ and $q( {%
\omega },1) $, we define 
\begin{equation}
\epsilon _{{\omega }}:=\left[ q( {\omega },0) -q( {\omega },1) \right] ^{2},
\end{equation}%
$\epsilon _{{\omega }}=0$ (1) for a satisfied (frustrated) wire.\ \ Then the
error metric is 
\begin{equation}
\epsilon _{w}:=\sum_{{\omega }=1}^{W}\epsilon _{{\omega }},
\end{equation}%
the number of frustrated wires. We do not engineer the error metric operator
here. Solving the given problem requires minimizing the wire error subject
to the triode relations.

Each triode is associated with three orthogonal eigenmodes $|\theta \rangle $
defined so that the binary variable $s_{\theta }^{2}$ has the value 1 for
the mode $|\theta \rangle $ and 0 in the other two modes.

The three-dimensional Hilbert space of the triplet modes of triode $\tau $
we designate by $\mathcal{H} _{\tau }$. We define an auxiliary mode space of 
${T}$ disconnected triodes as the tensor product $\bigotimes_{\tau =1}^{T}\ 
\mathcal{H} _{\tau }\equiv \mathcal{H} ^{T}$, a Hilbert space of dimension $%
3^{T}$.

The total error form $\epsilon_{w}$ of the network is a lower bound on the
number of wires that have to be changed to attain the solution. The number
could be as great as ${W}$ even if $\epsilon _{w}=1$.

We turn the error function into the Hamiltonian $H_{w}=g\epsilon _{w}$
diagonal in the qubit basis, with a coefficient $g$ to provide the dimension
of energy.

\section{ Computation model}

For qusa computation we take the effective Hamiltonian $H_{w}+H_{r}(t)$,
where $H_{r}(t)$ is a small effective\ Hamiltonian term (possibly
non-Hermitian) representing relaxation processes that bring the network to
the ground mode of $H_{w}$. $H_{r}(t)$ is discussed in the following.

\subsection{Continuous statistical projection}

\label{sec:SYM}

To estimate qusa speed-up, we represent a development subject to statistics
as a continuous projection of a dynamical development not subject to
statistics.

Consider a pair of identical protons 1 and 2. We freeze their spatial mode
to a fixed antisymmetric wave-function $\psi _{12}(x_{1},x_{2})$ so that
only the spin degrees of freedom need be considered. The individual spin
modes form two-dimensional Hilbert spaces $\mathcal{H}_{1}$ and $\mathcal{H}%
_{2}$. We consider:

\begin{itemize}
\item an unsymmetrized tensor-product (``comparison'') Hilbert space $%
\mathcal{H}^{\prime }=\mathcal{H}_{1}\otimes \mathcal{H}_{2}=\mathcal{H}%
_{t}\oplus \mathcal{H}_{s}$, the direct sum of the physical triplet subspace 
$\mathcal{H\equiv H}_{t}$ and the singlet subspace $\mathcal{H}_{s}$,

\item a particle exchange operator $X_{12}:\mathcal{H^{\prime }}\rightarrow 
\mathcal{H}^{\prime }$,

\item a projection $P_{12}:=\frac{1}{2}(1+X_{12})$ on the physical triplet
subspace\ $\mathcal{H}\equiv \mathcal{H}_{t}\subset \mathcal{H}^{\prime }$

\item a symmetrized Hamiltonian $H_{12}:\mathcal{H}_{t}\rightarrow \mathcal{H%
}_{t}$ invariant under proton exchange: $X_{12}H_{12}=H_{12}X_{12},$ so that 
$X_{12} $ is a constant of the motion.

\item the extension $H$ of the spin Hamiltonian operator $H_{12}$ from the
triplet subspace $\mathcal{H}_{t}\subset \mathcal{H}^{\prime }$ to the
entire space $\mathcal{H}^{\prime },$ with $H=0$ on the singlet space for
convenience.
\end{itemize}

The orthohydrogen modes are symmetric in the spin variables. They vary over
the effective three-dimensional Hilbert space of triplet modes $\mathcal{H}%
_{t}\subset \mathcal{H}^{\prime}.$

In the Hilbert space $\mathcal{H}^{\prime }$, the time development of the
orthohydrogen spins is then governed by a Hamiltonian $H$ that maps the
triplet subspace into itself: 
\begin{equation}  \label{eq:T}
\left| dt\right\rangle :=\left( 1-iHdt\right) \left| 0\right\rangle \/,
\end{equation}
where $\left| 0\right\rangle $\ is a symmetric mode in $\mathcal{H}^{\prime
}\ . $ We now develop an equivalent representation of the time development.
The aim is to free the Hamiltonian from the mathematical conditions
representing proton indistinguishability.

We start from a symmetric initial mode $\left| 0\right\rangle $ of $\mathcal{%
H}^{\prime }$ at time 0 and let it evolve for an infinitesimal amount of
time $dt$ according to a different comparison Hamiltonian $H^{\prime }$:

\begin{equation}  \label{eq:T'}
\left| dt\right\rangle ^{\prime }:=\left( 1-iH^{\prime }dt\right) \left|
0\right\rangle .
\end{equation}%
$H^{\prime }$ is not subject to exchange symmetry, but we choose it so that
its symmetrization yields the actual Hamiltonian: 
\begin{equation}
P_{12}H^{\prime }P_{12}=H.
\end{equation}

In general $\left| dt\right\rangle ^{\prime }$ is not symmetric under proton
exchange. We restore particle indistinguishability by projecting $\left|
dt\right\rangle ^{\prime }$ on $\mathcal{H}_{t}$. This means symmetrizing $%
\left| dt\right\rangle ^{\prime }$ to form 
\begin{equation}
P_{12}\left| dt\right\rangle ^{\prime }\equiv \left( 1-iHdt\right) \left|
0\right\rangle {}.
\end{equation}%
The projection of the infinitesimal development (10) on $\mathcal{H}_{t}$
yields the actual development (9), up to higher order infinitesimals. We
``continuously project'' on $\mathcal{H}_{t}$ the development governed by $%
H^{\prime }$. That is, we project after each interval $\Delta t$ and take
the limit $\Delta t\rightarrow 0$. This recovers the actual development
generated by the symmetrized Hamiltonian $H$.

This mathematical artifice of asymmetric time-development accompanied with
continuous symmetrization permits us to estimate the speed-up due to quantum
statistical computation.

\subsection{The speed-up due to statistics}

For qusa computation, we apply the continuous symmetrization of Section \ref%
{sec:SYM} to the relaxation of the triode network; see for example (\ref%
{eq:TOY}). For a comparison network we work in the unsymmetrized tensor
product Hilbert space $\mathcal{H}^{\prime }\supset \mathcal{H}^{T}$,
suspending proton indistinguishability and removing all the consequent
statistical relations.

This means dropping the triode condition $q_{x}+q_{y}+q_{z}=1$ for the
weaker condition $q_{x}+q_{y}+q_{z}=1$ mod 2. The latter holds independently
of statistics due to the composition of angular momentum alone, so it
survives.

$H_{w}$ usually has traps (local minima relative to all immediately adjacent
energy levels) that slow classical annealing computation. We eliminate these
for the comparison computation by redefining $H_{w}$ in a way that does not
change the ground mode: 
\begin{equation}
H_{w}=g\epsilon _{w}+g^{\prime }\epsilon _{w}\sum_{\tau }\left\{ \left[ 1-q(
\tau ,1) \right] ^{2}+\left[ 1-q( \tau ,2) \right] ^{2}+\left[ 1-q\left(
\tau ,3\right) \right] ^{2}\right\} ,
\end{equation}%
where $q( \tau ,1) ,...$ are the three qubits of triode $\tau $.

Since for the triode network each triode has only one qubit equal to $1$, we
have merely multiplied the previous Hamiltonian by $\left( 1+2T\frac{%
g^{\prime }}{g}\right) .$ This does not change the ground mode.

In the case of the EQU network, all the three qubits of a triode can be $1.$
If $g^{\prime }\gg g$, each frustrated network mode ($\epsilon _{w}\geq 1$)
has a gradient toward the solution where all qubits are $1$. This ground
mode is quickly reachable by the EQU network even in classical annealing.

We define the effective Hamiltonians of the actual and comparison networks, 
\begin{eqnarray}
H &=&H_{w}+H_{r}, \\
H^{\prime } &=&H_{w}^{\prime }+H_{r}^{\prime }.
\end{eqnarray}%
Each describes a network with symmetric wire Hamiltonian $H_{w}$ or $%
H_{w}^{\prime }=H_{w}$, coupled to a heat bath by an effective coupling
Hamiltonian respectively $H_{r}$ and $H_{r}^{\prime }$. $H_{r}^{\prime }$ is
not symmetrized; its symmetrization yields $H_{r}$. $H$ describes the
relaxation process of the actual triode network. $H^{\prime }$ describes the
relaxation process of a comparison network obtained by replacing all pairs
of identical indistinguishable spin $1/2$ particles with \ pairs of
distinguishable (e.g. non-identical) spin $1/2$ particles. Correspondingly,
all triode relations are replaced by EQU gate relations.

We model the actual heat bath coupling $H_{r}$ by coupling each proton spin $%
\vec{\sigma}$ to a small Gaussian random time-varying magnetic field $\vec{B}
$ at the site of that spin. 
$\vec{B}$ might be polarized along the principal
direction $x+y+z$. We index the sites with the triode index $\tau =1,\dots ,{%
T}$ and a binary index $\beta =1,2$. While $\tau $ enumerates the triodes
(proton pairs), $\beta $ distinguishes the two protons in each triode.

To preserve statistics we must demand that the two protons $\beta =1,2$ of
each triode $\tau $ experience the same magnetic field $\vec{B}(\tau )$. We
may then write the actual relaxing coupling as 
\begin{equation}
H_{r}=g\sum_{\tau ,\beta }\vec{B}(\tau )\cdot \vec{\sigma}(\tau ,\beta )
\end{equation}

The comparison heat bath is a random magnetic field at each proton site.
Unlike the actual heat bath coupling, the comparison heat bath coupling ${H}%
_{r}^{\prime }$ is not invariant under proton exchange. Different protons in
the same triode see different magnetic fields $\vec{B}(\tau ,\beta, )$: 
\begin{equation}
{H}_{r}^{\prime }=g\sum_{\tau ,\beta }\vec{B}(\tau ,\beta )\cdot \vec{\sigma}%
(\tau ,\beta )
\end{equation}%
$\vec{B}(\tau ,\beta )$ too might be polarized along 
the principal direction
$x+y+z$.

Let $P$ be the symmetrization operator for all the relevant proton
permutations; it is not necessary to permute protons between triodes. We may
arrange that the projected heat-bath coupling is the actual coupling,

\begin{equation}
PH_{r}^{\prime }P=H_{r},
\end{equation}%
by identifying the random magnetic field $\vec{B}(\tau )$ of the actual heat
bath with the average of the two random magnetic fields of the comparison
heat bath: 
\begin{equation}
\vec{B}(\tau )\equiv {\frac{\vec{B}(\tau ,1)+\vec{B}(\tau ,2)}{2}}
\end{equation}%
The sum of two Gaussian random variables is also a Gaussian random variable.

Summing up, we have $PH_{w}^{\prime }P=H_{w},$ $PH_{r}^{\prime }P=H_{r},$
and thus $PH^{\prime }P=H$. Therefore we can apply the method of Section \ref%
{sec:SYM}.

The fact that $H_{w}^{\prime }=H_{w}$ is already symmetric with respect to
proton exchange (as all its qubits are) does not introduce any unwanted
constraint in the comparison relaxation process. It does not prevent the
generation of triode violations by $H_{r}^{\prime }$. $H_{w}$ symmetry only
reflects the composition of angular momentum, not statistics.

Here we do not tackle the difficult problem of estimating the relaxation
time of the actual triode network, or the comparison EQU network. We just
know that $H^{\prime }$\ will eventually drive \ the EQU network to its
ground mode, and compare the relaxation times of the two networks.

For this comparison, decompose the actual computation time $\Delta T$ into $%
N=\frac{\Delta T}{\Delta t}$ consecutive time slices $\Delta t_{i}:t_{i}\leq
t\leq t_{i+1}$ of equal length $\Delta t$, with $i=1,2,...,N$, $%
t_{i}=i\Delta t$. Take the relaxation within each $\Delta t_{i}$ to be that
of the comparison EQU network. At the end of each $\Delta t_{i}$ project the
network mode on the Hilbert space $\mathcal{H}^{T}.$ Then take the limit $%
\Delta t\rightarrow 0$. This reproduces the actual computation.

Let $\left| 0\right\rangle $ be an initial symmetrical preparation, where
all triodes must be satisfied and wires can be frustrated. Let this evolve
into $\left| t\right\rangle $ at time $t$, with random phases as required.
To consider the development of $\left| t\right\rangle $ inside the interval $%
\Delta t_{i}$, we resolve $\left| t\right\rangle $ as follows:%
\begin{equation}
\left| t\right\rangle =\left| S,t\right\rangle +\left| F,t\right\rangle
+\left| V,t\right\rangle .
\end{equation}%
$\left| S,t\right\rangle $ denotes a superposition of tensor products of
triode eigenmodes (terms) which are solutions of the triode network (each
with a random phase to represent incoherence as necessary); its terms have
satisfied triodes and satisfied wires. Most probably, relaxation randomly
generates a solution with probability $p_{S}(t):=\langle S,t|S,t\rangle $ of
the order of $1/2^{Q}$ in poly$(Q)$ time. We assume this is the case at time 
$t_{i}$. Since 
\begin{equation}
H_{w}\left| S,t\right\rangle =0,
\end{equation}%
we can assume that $\left\langle S,t|S,t\right\rangle $ remains
approximately constant inside $\Delta t_{i}.$

$\left| F,t\right\rangle $ is the component of $|t\rangle $ whose terms have
satisfied triodes and at least one frustrated wire. Its probability $%
p_{F}(t):=\left\langle F,t|F,t\right\rangle $ is initially close to $1$ (see
further below).

$\left| V,t\right\rangle $ is the component whose terms have at least one
violated triode; wires can be either satisfied or frustrated. Its
probability is $p_{V}(t):=\langle V,t|V,t\rangle $\/.

$\left| V,t\right\rangle $ is generated by the relaxation of the EQU network
inside each $\Delta t_{i}$; it goes to zero with $\Delta t$ and is
annihilated by the projection at the end of $\Delta t_{i}.$

We compare the rate of relaxation of the triode network to that of the EQU
network as follows.

Inside $\Delta t_{i}$, the evolution is that of the comparison network where
all triodes are replaced by EQU gates. Therefore $p_{F}(t)$ goes down at the
rate of relaxation of the EQU network, building up the mode $\left|
V,t\right\rangle $.

We are particularly interested in the ``take off'' of the solution
probability $p_{S}(t)$ from $O\left( 1/2^{Q}\right) $ to \ $O(1)$, say to $%
p_{S}(t)=1/10$. During take off, we can assume a constant rate of relaxation 
$k$ of the EQU network with $p_{F}(t)\approx 1$. That is, $p_{F}(t)$ changes
from $p_{F}(t_{i})\approx 1$ at the beginning of $\Delta t_{i}$ to 
\[
p_{F}(t_{i+1})\approx \left( 1-k\Delta t\right) p_{F}(t_{i})\approx \left(
1-k\Delta t\right) 
\]%
at the end of $\Delta t_{i}$. Correspondingly, $p_{V}(t)$ changes from $0$
to about $k\Delta t$.

The projection at $t_{i}+\Delta t_{i}$ therefore reduces $\langle t|t\rangle$
by about $k\Delta t$. Renormalizing then multiplies \ $p_{S}(t)$ by about $%
(1-k\Delta t)^{-1}\approx 1+k\Delta t$ at each $\Delta t_{i}$. After a time $%
N\Delta t=$ $\Delta T$ we have 
\begin{equation}
p_{S}(t_{i}+\Delta T)=p_{S}(t_{i})\lim_{\Delta t\rightarrow 0}\left(
1+k\Delta t\right) ^{\frac{\Delta T}{\Delta t}}=p_{S}(t_{i})e^{k\Delta T}.
\end{equation}

Thus $p_{S}(t)$ becomes $O(1)$ in a time $\Delta T$ such that $e^{k\Delta
T}=O(2^{Q}).$ This means $k\Delta T=O(Q)$, the number of qubits. The
relaxation time\ $\Delta T$ of the actual network is comparable with that of
the easy EQU network.

Let us check that assuming a constant (average) relaxation rate does not
introduce unwanted restrictions. We show that the continuous projection
method works in the same way in presence of fluctuations of the expected
energy of the EQU network. Assume this energy goes up in the time interval $%
\left[ t_{i},t_{j}\right] ,$ with $t_{j}>t_{i}$, namely $\left\langle
t_{j}\right| H_{w}\left| t_{j}\right\rangle >\left\langle t_{i}\right|
H_{w}\left| t_{i}\right\rangle .$ On the basis of (22), this brings $%
p_{S}(t) $ down but, as readily checked, what is lost is exactly regained
when the expected energy goes back to $\left\langle t_{i}\right| H_{w}\left|
t_{i}\right\rangle $ at some time $t_{h}>t_{j}.$

Summing up: (i) We have defined an effective symmetrized Hamiltonian $H$ for
the relaxation of the ``hard'' triode network. (ii) We have defined a
comparison non-symmetrical Hamiltonian $H^{\prime }$ whose symmetrization
yields $H$: $PH^{\prime }P=H$. (iii) Furthermore $H^{\prime }$ describes the
relaxation process of an easy EQU network, obtained by replacing all triode
relations by EQU relations. (iv) The relaxation of the hard triode network
can be obtained by continuously projecting that of the easy EQU network on
the symmetrical subspace $\mathcal{H}^{T}$. By eq. (22), this yields
comparable relaxation times for the hard and easy networks.

Satisfying relations implemented by particle statistics does not take time
in annealing computation.

If quantum annealing can reach the solution $111...$ of a general EQU
network in polynomial time, then in qusa computation NP=P.

Qusa computation survives decoherence as well as general annealing
computation does. They both avoid this basic difficulty of reversible
quantum computation.

\section{Discussion}

\subsection{The origin of the qusa speed-up}

What makes quantum computation more efficient than classical computation has
been called a ``most pressing'' question for the advancement of the field
(Mahler 2001). The speed-up of the quantum algorithms (Deutsch 1985, Shor
1994, Grover 1996, among others) stems from the fact that a quantum
transition is jointly influenced by an initial and a final extra-dynamical
selection. It may be regarded as extra-dynamical in origin (Castagnoli \&
Finkelstein 2001). This kind of process is richer than classical
computation, which is a dynamical development of one selection alone.

Qusa computation can be seen as a dynamical development generated by a
symmetric time-varying Hamiltonian $H$ with a time-dependent magnetic field $%
\overrightarrow{B}$ (varying randomly for annealing). The extra-dynamical
origin of its speed-up is $H$ symmetry itself. This originates
extra-dynamically in proton indistinguishability, not dynamically. By the
way, one can see from equations (10) and (12) that symmetrizing the network
mode and symmetrizing $H^{\prime }$ are equivalent.

It is as if proton indistinguishability provides for free an extra-dynamical
symmetrization engine continously acting on an un-symmetrized $H^{\prime }$
or the development generated by it (relating the easy EQU network). The
comparison with the quantum algorithms is easier if we consider the
development.

Qusa speed-up comes from the fact that the Boolean statistical relations are
satisfied through the projection of a development that is unaffected by
them. Correspondingly, we have seen that in each $\Delta t_{i}$ there is a
quantum transition jointly influenced by an initial and a final
extra-dynamical selection. This is as in the quantum algorithms, but for the
fact that qusa computation always selects a predetermined subspace, the one
that satisfies statistics.

\subsection{Conclusions}

Qusa speeds quantum ground mode computation up by implementing the gates (or
partial gates) of a Boolean network with statistical symmetries, and only
the wires through an energy function which is zero when the wire is
satisfied. Logical relations associated with statistical symmetries do not
slow down the development of the initial mode toward a ground mode where
both gates and wires are satisfied. This is unlike classical ground mode
computation, where logical relations reduce relaxation rate.

Qusa computation develops quantum parallelism through the incoherent
superposition of parallel computation paths (i.e. through mixtures). It
abandons the delicate superposition of coherent parallel computation paths
in reversible quantum computation but introduces the almost indestructible
superposition of different permutations of identical particles subject to a
given statistics. This greatly reduces the problem of decoherence.

This synthesis of \ ground mode computation and quantum statistics appears
to be a promising architecture for robust quantum computing.

A natural next step in this research is to design physical systems
exhibiting the qusa effect.

\bigskip\ 

\textbf{ACKNOWLEDGMENTS} \bigskip

The ideas propounded in this work were developed through discussions with
Artur Ekert. We thank Shlomit Ritz Finkelstein for useful discussions of the
presentation, Lou Pagnucco for kindly pointing out an error that we have
corrected, and William M. Kaminsky for helpful remarks.

\bigskip

\textbf{REFERENCES}

\begin{enumerate}
\item Castagnoli, G. 1998 \emph{Physica} D \textbf{120}, 48.

\item Castagnoli, G., Ekert, A. \& Macchiavello, C. 1998 \emph{Int. J.
Theor. Phys.} \textbf{37}, 463.

\item Castagnoli, G. \& Monti, D. 1999 \emph{Chaos, Solitons \& Fractals} 
\textbf{10}, 1665.

\item Castagnoli, G. \& Finkelstein, D.R. 2001 \emph{Proc. R. Soc. Lond.} A 
\textbf{457}, 1799.

\item Deutsch, D. 1985 \emph{Proc. R. Soc. Lond.} A \textbf{400}, 97.

\item Farhi, E., Goldstone, J., Gutmann, S., Lapan, J., Lundgren, A. \&
Preda, D. 2001 \emph{Science} \textbf{292}, 472.

\item Finkelstein, D. 1996 \emph{Quantum Relativity}\/. Springer.

\item Grover, L. 1996 In \emph{Proc. 28th A. ACM Symp. on Theory of Computing%
}, P. 212. Philadelphia, PA: ACM Press.

\item Kirkpatrick, S. \& Selman, D. 1994 \emph{Science} \textbf{264}, 1297.

\item Mahler, G. 2001 \emph{Science} \textbf{292}, 57.

\item Shor, P. 1994 In \emph{Proc. 35th A. Symp. of the Foundation of
Computer Science, Los Alamitos, CA}, P. 124. Los Alamitos, CA: IEEE Computer
Society Press.
\end{enumerate}

\bigskip

\end{document}